%% file: main.tex
\documentclass[sigconf,10pt]{acmart}
\renewcommand\footnotetextcopyrightpermission[1]{} 
\setcopyright{none}
\usepackage[small, compact]{titlesec}

\acmDOI{}

\acmISBN{}

\acmPrice{}

\PassOptionsToPackage{pdfpagelabels=false}{hyperref}

\usepackage{amsmath}
\usepackage{gnuplottex}
\usepackage{xspace}
\usepackage{subcaption}
\usepackage{caption}
\usepackage{booktabs}
\usepackage{float}
\usepackage{multicol}
\usepackage{multirow}
\usepackage{trig}

\newcommand{\captionfonts}{\bf \small}

\makeatletter  
\long\def\@makecaption#1#2{%
	\vskip\abovecaptionskip
	\sbox\@tempboxa{{\captionfonts #1: #2}}%
	\ifdim \wd\@tempboxa >\hsize
	{\captionfonts #1: #2\par}
	\else
	\hbox to\hsize{\hfil\box\@tempboxa\hfil}%
	\fi
	\vskip\belowcaptionskip}
\makeatother 

\newcommand{\squishlist}{
  \begin{list}{$\bullet$}{
    \setlength{\itemsep}{0pt}       \setlength{\parsep}{3pt}
    \setlength{\topsep}{3pt}        \setlength{\partopsep}{0pt}
    \setlength{\leftmargin}{1em}    \setlength{\labelwidth}{1em}
    \setlength{\labelsep}{0.5em} } }

\newcommand{\squishend}{
  \end{list} }

\def\compactify{\itemsep=2pt \topsep=2pt \partopsep=1pt \parsep=1pt \leftmargin=1.2em}
\let\latexusecounter=\usecounter

\usepackage[textsize=tiny,textwidth=0.6in]{todonotes}
\newcommand{\allnotes}[1]{}
\renewcommand{\allnotes}[1]{#1} 

\newcommand{\subdraft}[1]{}

\newcommand{\tianyi}[1]{\allnotes{\todo[color=green!50]{TC: #1}}}
\newcommand{\wei}[1]{\allnotes{\todo[color=blue!50]{Wei: #1}}}

\newcommand{\kaiyuan}[1]{\allnotes{\todo[color=gray!30]{Kaiyuan: #1}}}

\newcommand{\sys}{Laconic\xspace}

\begin{document}

\title{\Huge{Laconic: Streamlined Load Balancers for SmartNICs}}
\author{Tianyi Cui}
\affiliation{
\institution{University of Washington}
\country{}}
\email{cuity@cs.washington.edu}
\author{Chenxingyu Zhao}
\affiliation{
\institution{University of Washington}
\country{}}
\email{cxyzhao@cs.washington.edu}
\author{Wei Zhang}
\affiliation{
\institution{Microsoft}
\country{}}
\email{wei.zhang.gbs@gmail.com}
\author{Kaiyuan Zhang}
\affiliation{
\institution{University of Washington}
\country{}}
\email{kaiyuanz@cs.washington.edu}
\author{Arvind Krishnamurthy}
\affiliation{
\institution{University of Washington}
\country{}}
\email{arvind@cs.washington.edu}

\begin{abstract}
\input{sections/abstract}
\end{abstract}
\maketitle
\pagestyle{plain}

\input{sections/intro}
\input{sections/background}

\input{sections/design}
\input{sections/eval_v2.tex}

\input{sections/relwk}
\input{sections/conclusion}


\normalsize 
\bibliographystyle{plain}
\bibliography{reference}

\clearpage
\appendix
\begin{sloppypar}
\input{sections/appendix.tex}
\end{sloppypar}


\end{document}

%% file: sections/abstract.tex
Load balancers are pervasively used inside today's clouds to scalably distribute network requests across data center servers.  Given the extensive use of load balancers and their associated operating costs, several efforts have focused on improving their efficiency by implementing Layer-4 load-balancing logic within the kernel or using hardware acceleration.  This work explores whether the more complex and connection-oriented Layer-7 load-balancing capability can also benefit from hardware acceleration.  In particular, we target the offloading of load-balancing capability onto programmable SmartNICs. We fully leverage the cost and energy efficiency of SmartNICs using three key ideas.  First, we argue that a full and complex TCP/IP stack is not required for Layer-7 load balancers and instead propose a lightweight forwarding agent on the SmartNIC.  Second, we develop connection management data structures with a high degree of concurrency with minimal synchronization when executed on multi-core SmartNICs. Finally, we describe how the load-balancing logic could be accelerated using custom packet-processing accelerators on SmartNICs. We prototype Laconic on two types of SmartNIC hardware, achieving over 150 Gbps throughput using all cores on BlueField-2, while a single SmartNIC core achieves 8.7x higher throughput and comparable latency to Nginx on a single x86 core.


%% file: sections/intro.tex
\section{Introduction}
\label{sec:intro}

Load balancers are a fundamental building block for data centers as they balance the service load across collections of application servers \cite{zhang2017resilient,tiara,miao2017silkroad}.  Load balancers were initially built as specialized hardware appliances but are now typically deployed as software running on commodity servers or VMs.  This deployment model provides a greater degree of customizability and adaptability than the older hardware-based designs, but it also can result in high costs for cloud providers and application services, given the purchase costs and the energy consumption of general-purpose servers~\cite{microserviceperf}. Application services often go to great lengths to consolidate and reduce their use of load balancers to obtain desired cost savings~\cite{aws-lb,aws-elb,aws-lb-opts}.

Given the extensive use and cost of load balancers, several efforts have focused on improving their efficiency, especially Layer-4 (L4) load balancers, by embedding the load balancing logic in a lower, possibly hardware-accelerated, layer. Katran~\cite{katran} is accelerated using eBPF code inside the Linux kernel, thus intercepting and processing packets within the kernel and minimizing the number of transitions to user-level load-balancing code. ClickNP~\cite{clicknp} tackles some aspects of the L4 load balancing logic (especially NAT-like capabilities) on an FPGA-enabled SmartNIC and exploits the parallel processing capabilities of FPGA devices. SilkRoad~\cite{miao2017silkroad} uses a combination of a programmable switch and an end-host to store the state associated with L4 load balancers and perform the dataplane transformations related to the load-balancing operation within a switch pipeline. 

While these efforts have made considerable gains in optimizing L4 load balancing that balances traffic at the network layer, data center services often rely on application-layer load-balancing capabilities found only in Layer-7 (L7) load balancers. In particular, services would like to route flows based on the attributes of the client request, preserve session affinity for client requests, provide access control, and so on~\cite{aws-alb}. But, these features make it harder for L7 load balancers to adopt the hardware-acceleration techniques used for L4 load balancers. A fundamental challenge is that the L7 load-balancing operation is based on information embedded within connection-oriented transport protocols, thus seeming to require a full-stack network processing agent on the load balancer to handle TCP/HTTP connections. Consequently, today's L7 load balancers are generic software solutions incurring high processing costs on commodity servers.

In this work, we examine whether we can improve the efficiency of L7 load balancers using programmable networking hardware. We focus on SmartNICs that provide general-purpose computing cores augmented with packet-processing hardware. SmartNICs are particularly attractive targets as their compute cores can host arbitrary protocol logic while their packet-processing accelerators can perform dataplane transformations efficiently. A SmartNIC thus combines the capabilities of traditional host computing with the emerging capabilities of programmable switches and is an appropriate target for L7 load balancers. Our work is also partly motivated by the increasing deployment of SmartNICs within data centers as a cost-effective and energy-efficient computing substrate for networking tasks. 



Several challenges have to be addressed in offloading load-balancing functionality to SmartNICs. First, SmartNIC cores are wimpy, equipped with limited memory, and inefficient at running general-purpose computations.  To the extent possible, we should use lightweight network stacks instead of generic, full-functionality stacks present inside OS kernels.  Second, efficient multi-core processing on the SmartNICs presumes lightweight synchronization for access to concurrent data structures, and this is particularly relevant as we slim down the network processing logic.  Third, the effective use of accelerators for packet transformations is necessary to enhance the computing capability of SmartNICs.


We design and implement \textit{\sys}, a SmartNIC-offloaded load balancer that addresses the challenges raised above.  A key component of our system is a lightweight network stack that represents a co-design of the application-layer load-balancing logic with the transport layer tasks. This lightweight network stack performs complex packet processing only on a subset of the packets transmitted through the load balancer. For the rest of the packets, the network stack performs simple rewriting of packets and relies on the client and the server to provide end-to-end reliability and congestion control. For SmartNICs with packet-processing accelerators, this design allows most of the packets to be processed using hardware-based flow-processing engines, thus providing significant efficiency gains. 
We also develop connection management data structures that are highly concurrent and minimize expensive mutual exclusion operations.  
We note that some of our design contributions also apply to a generic server-based design, but they have a multiplicative effect on SmartNICs by factoring out a fast path that can be executed on the hardware packet engines.


We built Laconic and tailored it to two different types of SmartNICs: Marvell LiquidIO3 and Nvidia BlueField-2. Laconic provides both Layer-4 and Layer-7 functionality and implements commonly used Layer-7 interposition logic for balancing connections to backend services. 
For large messages, \sys running on BlueField-2 with one single ARM core can achieve up to 8.7x higher throughput than widely-used Nginx running on a more powerful x86 core. For small messages, \sys on BlueField-2 can achieve higher throughput with comparable or even lower latency compared with Nginx. On LiquidIO3, the throughput of Laconic is 4.5x higher compared with x86 Nginx. We also demonstrate the Laconic performance with real-world workload and present detailed microbenchmarks on the benefits of key ideas.



%% file: sections/background.tex
\section{Background}

This section discusses the structure, connectivity, and performance of programmable multi-core SmartNICs that are commonly available. We also describe the load-balancing functionality we want to deploy on these SmartNICs.

\subsection{Programmable Multi-core SmartNICs}
\label{subsec:smartnic}

We consider programmable SmartNICs equipped with multi-core processors.  A typical SmartNIC has onboard memory, DMA engines, and accelerators (e.g., engines for crypto, compression, and packet rewriting) in addition to the multi-core processor.  Below, we discuss the two dominant categories, on-path and off-path SmartNICs~\cite{liu2019offloading}. 

\begin{figure}[tbp]
\centering
\includegraphics[width=0.4\textwidth]{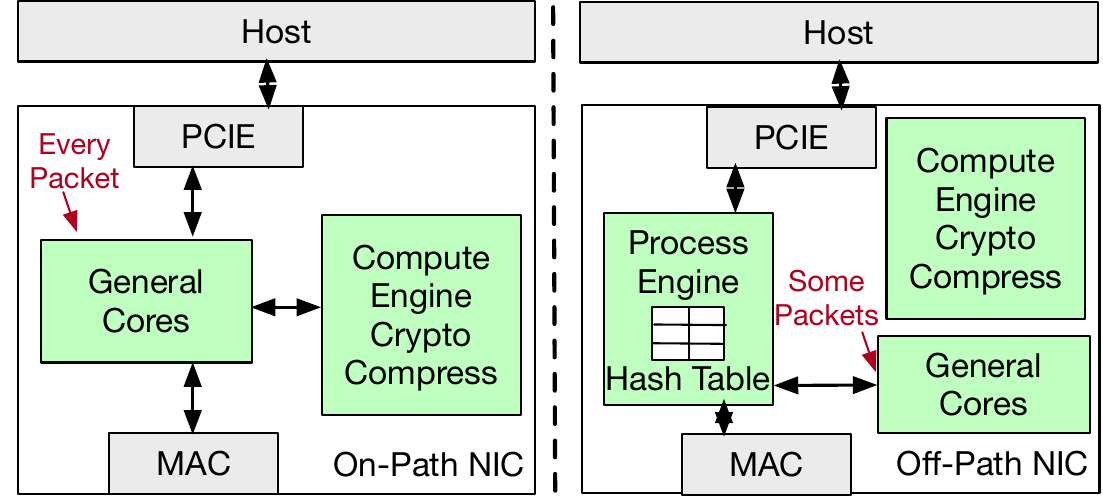}
\caption{Common SmartNIC architectures}
\label{fig:smartNIC-arch}
\end{figure}

\vspace{0.1in}
\noindent \textbf{On-Path SmartNICs:} These are SmartNICs where the NIC processing cores are on the data path between the network port and the host processor (see Figure~\ref{fig:smartNIC-arch}).  Consequently, every packet received or transmitted by the host is also processed by the NIC cores.  The performance of the NIC cores is critical to the throughput and latency characteristics of the NIC.  To address this issue, these NICs typically augment the traditionally wimpy cores on the SmartNIC with specialized hardware support that enhances the packet-processing capabilities of the core. For example, packet contents are prefetched and placed in a structure similar to the L1 cache, and there are hardware mechanisms for managing packet buffers. The SmartNIC has hardware mechanisms for delivering incoming packets to NIC cores in a balanced manner, but there are no mechanisms, such as receive side scaling (RSS), to deliver specific packets to specific NIC cores.  Further, the NIC cores can invoke specialized accelerators for tasks such as crypto and compression.  Marvell LiquidIO~\cite{liquidio} and Netronome NICs~\cite{agilio} are on-path SmartNICs.

\noindent \textbf{Off-Path SmartNICs:} These are SmartNICs where the NIC's processing cores are off the data path connecting the host to the network. A NIC-level switching fabric, a NIC switch, provides connectivity between the network port, the host cores, and the NIC cores. The NIC switch is a specialized hardware unit with match-action engines for selecting packet fields and rewriting them based on runtime-configurable rules.  The rules route packets received from the network to the host or the NIC, or rewrite them and immediately transmit them back into the network.  Mellanox Bluefield~\cite{bluefield} and Broadcom Stingray~\cite{stingray} are off-path SmartNICs.

The off-path SmartNICs thus contain both general-purpose cores and packet match-action engines. The packet-processing logic could thus be split across the cores and the match-action engines based on the complexity of the logic. For instance, the cores can perform complex packet processing and delegate to the match-action engines simpler operations such as packet steering, header-rewriting, and packet forwarding.  


\vspace{0.1in}
\noindent \textbf{Offloading to SmartNICs:} Datacenter operators are increasingly resorting to offloading various infrastructure functions to SmartNICs and reducing the usage of host cores, which can, in turn, then be rented out and monetized. The ARM cores on SmartNICs are not as powerful as the host x86 cores, but they are an order of magnitude less expensive and result in substantial cost savings~\cite{fungible-pr,netronome-pr}.
Network virtualization, security, and storage are some of the typical operations offloaded to SmartNICs, and in this work, we seek to expand this to include the infrastructure's load-balancing function. 

We seek to offload our load-balancing capability on both on-path and off-path load balancers. Others have noted that the communication latencies between the SmartNIC cores and the host cores can be significant in the case of off-path SmartNICs~\cite{liu2019offloading}, so we target complete offloads for both types of SmartNICs, i.e., all load balancing logic is performed on the SmartNIC cores. Further, this allows us to target even headless SmartNICs with an independent power supply and a carrier card, allowing for a host-less solution (e.g., Broadcom Stingray PS1100R~\cite{ps1100r}) and enabling substantial reductions to both hardware and operating costs.

\subsection{Load balancers}
\label{subsec:backgroundlb}

Load balancers operating at different network layers are widely deployed inside data centers to deliver traffic to services. They fall into two categories: Layer 4 and Layer 7.

An L4 load balancer maps a virtual IP address (VIP) to a list of backend servers, with each server having its own dynamic IP address (DIP). As the L4 load balancers operate on the transport layer, the routing decision is solely based on the packet headers of the transport/IP layers (i.e., the 5-tuple of IP addresses and ports) but not the payload. Several L4 load balancers are deployed by cloud providers~\cite{patel2013ananta,eisenbud2016maglev,katran}).  


L7 load balancers are significantly more complex and operate on content at the application layer (e.g., HTTP data). Many services inside a data center may share a common application gateway implementing the L7 load-balancing functionality.  The L7 load balancer dispatches requests to the corresponding backend servers based on the service requested, e.g., different services are commonly differentiated by the URL~\cite{googleapigateway}. The load balancer would then reassemble the stream and match the URL against various patterns to route the request to the corresponding services.

It is common for an L7 load balancer to modify the streamed application data, e.g., insert an \texttt{x-forwarded-for} header to inform the backend server of the real IP address of the client. The load balancer may also modify the reply from the server to inject a cookie into the response. Future requests from the same client can then be routed to the same backend server based on the cookie.  


Due to the complex nature and the functionality provided by the L7 load balancer, it is typically implemented as a dedicated application on top of the kernel networking stack, e.g., Nginx~\cite{nginx}, Envoy~\cite{envoy}, HAProxy~\cite{haproxy}. Given the overheads of the OS's networking stack and the application-level load balancing, the performance of L7 load balancers is an order of magnitude lower than that of L4 load balancers, with 50\% to 90\% of the processing time spent inside the kernel~\cite{li2019socksdirect,jeong2014mtcp}.


%% file: sections/design.tex
\section{SmartNIC-based Load Balancers}

Laconic is a load balancer (LB) designed for SmartNICs that effectively uses their packet-processing capabilities.  Our work primarily targets L7 load balancers, but some techniques also apply to accelerating L4 load balancers as well as optimizing L7 load balancers running on traditional servers.

In this section, we first provide some baseline characterization experiments that empirically quantify the suitability of running a traditional load balancer on NIC cores, then provide an overview of the design of \sys\ before describing in detail the core techniques used by the system.

\subsection{Offloading to SmartNICs: Challenges and Opportunities}
\label{subsec:motive-exp}

We perform characterization experiments that quantify the cost of running an unmodified load balancer on the SmartNIC and then examine the performance of hardware acceleration features on SmartNICs. We demonstrate that while SmartNICs are not a good fit for directly porting LB software written for the host, customizing the software to leverage SmartNICs' hardware accelerators can yield significant performance gains.

We take a stock Nginx 1.20 load-balancer and run it on both host cores and SmartNIC cores. We use an x86 server with a Xeon Gold CPU and Mellanox ConnectX-5 NIC for the host experiments. We compared the host x86 performance to the LiquidIO3 CN3380 (with 24 ARM cores @2.2GHz) and BlueField-2 (with 8 ARM A72 cores) SmartNICs. We serve static files on the backend servers using a cache-enabled Nginx to ensure the bottleneck is on the load balancer, and we use \texttt{wrk}~\cite{wrk} to generate the workload.

\begin{table}
\small
\centering
\begin{tabular}{ccrrr}\toprule
\multicolumn{2}{c}{Response size} & LIO3 & BlueField-2 & x86 \\\midrule
\multirow{2}{*}{1KB}
& 1 Core & 0.17 & 0.12 & 0.31 \\
& 8 Core & 1.01 & 0.51 & 2.36 \\
\midrule
\multirow{2}{*}{1MB}
& 1 Core & 5.08 & 3.07 & 10.84 \\
& 8 Core & 4.92 & 10.47 & 40.86 \\
\bottomrule
\end{tabular}
\vspace{0.2cm}
\caption{Nginx Performance across different platforms (Gbps)}
\label{tbl:perf-nginx}
\end{table}

Table~\ref{tbl:perf-nginx} provides the throughput achieved by the different configurations, as we vary the number of cores on the process (i.e., x86 cores on the host and ARM cores on the SmartNICs) for two different request sizes (1KB and 1MB). When we consider single-core experiments, we observe that the x86 core is about 1.8x to 3.5x more effective than the ARM cores on the SmartNIC. As we scale the number of cores on the processor, we observe a disparity between 4x and 8x. In both SmartNICs, we observe significant scalability bottlenecks; the LiquidIO3 achieves a lower line rate with eight cores than one core while performing 1MB transfers. These experiments quantify the inefficiencies of simply running an unmodified load balancer on the SmartNIC cores, which aren't architected to run complex LB software.


\begin{figure}[htbp]
\centering
 \includegraphics[width=0.3\textwidth, keepaspectratio]{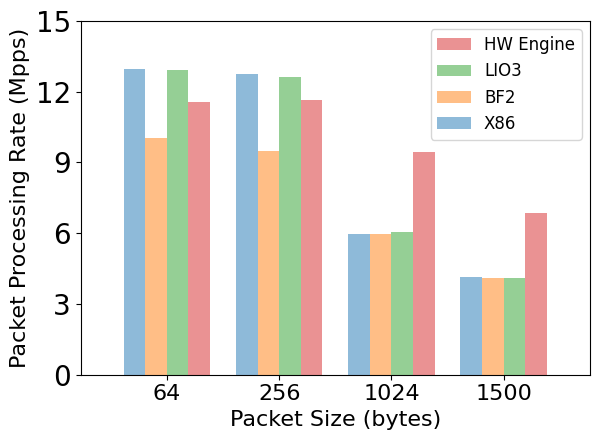}
\vspace{-0.2mm}
\caption{MAC-Swap performance across different platforms}
\label{fig:macswap-tput}
\end{figure}

We perform another characterization experiment related to basic packet forwarding capability. We measure the performance of a DPDK-based packet rewriting and forwarding program on all of our three platforms (i.e., host x86 cores, LiquidIO3 ARM cores, and BlueField-2 ARM cores). This benchmark simply forwards packets after swapping the MAC address fields on received packets. We consider different packet sizes and vary the number of operating cores (or queues). Figure~\ref{fig:macswap-tput} depicts the performance of our three platforms and the hardware flow processing engine of BlueField-2. In contrast to the Nginx benchmark, we observe that the disparity in packet processing performance between the three platforms is much lower, indicating that the SmartNIC cores are better suited for packet processing than the additional general-purpose computing required by Nginx.  Also, we demonstrate that the flow processing engine can operate efficiently compared with using x86 cores or ARM cores to process packets. Hardware flow processing engines can entirely offload packet processing and release the capability of the generic computing cores.  But, the hardware rules have limited processing capability, and there is a rule update cost (tens of $\mu s$, more details in Section \ref{sec:eval}) to inserting and deleting rules from the flow processing engine, which limits the performance gain for processing small messages with the hardware engine. Thus, we need a hybrid design combining generic NIC cores with a hardware flow processing engine. 

In summary, SmartNIC's ARM cores can match x86 cores for simple packet processing but are significantly worse for executing a full-featured network stack. The SmartNIC can, however, enjoy the benefits of hardware acceleration if the packet processing can be entirely performed using the flow processing engine.

\subsection{Laconic Overview}
\label{subsec:overview}

As observed above, unmodified L7 load balancers incur significant overheads when executed on the SmartNIC cores. The SmartNIC cores are significantly worse than the host cores for running a full-featured network stack that relies on the OS kernel to maintain channel state and provide reliable and sequenced delivery channels. We, therefore, develop a streamlined version of the load balancer that is explicitly tailored to the capabilities of SmartNICs. 

\begin{figure}[tbp]
\centering
\includegraphics[width=0.45\textwidth, keepaspectratio]{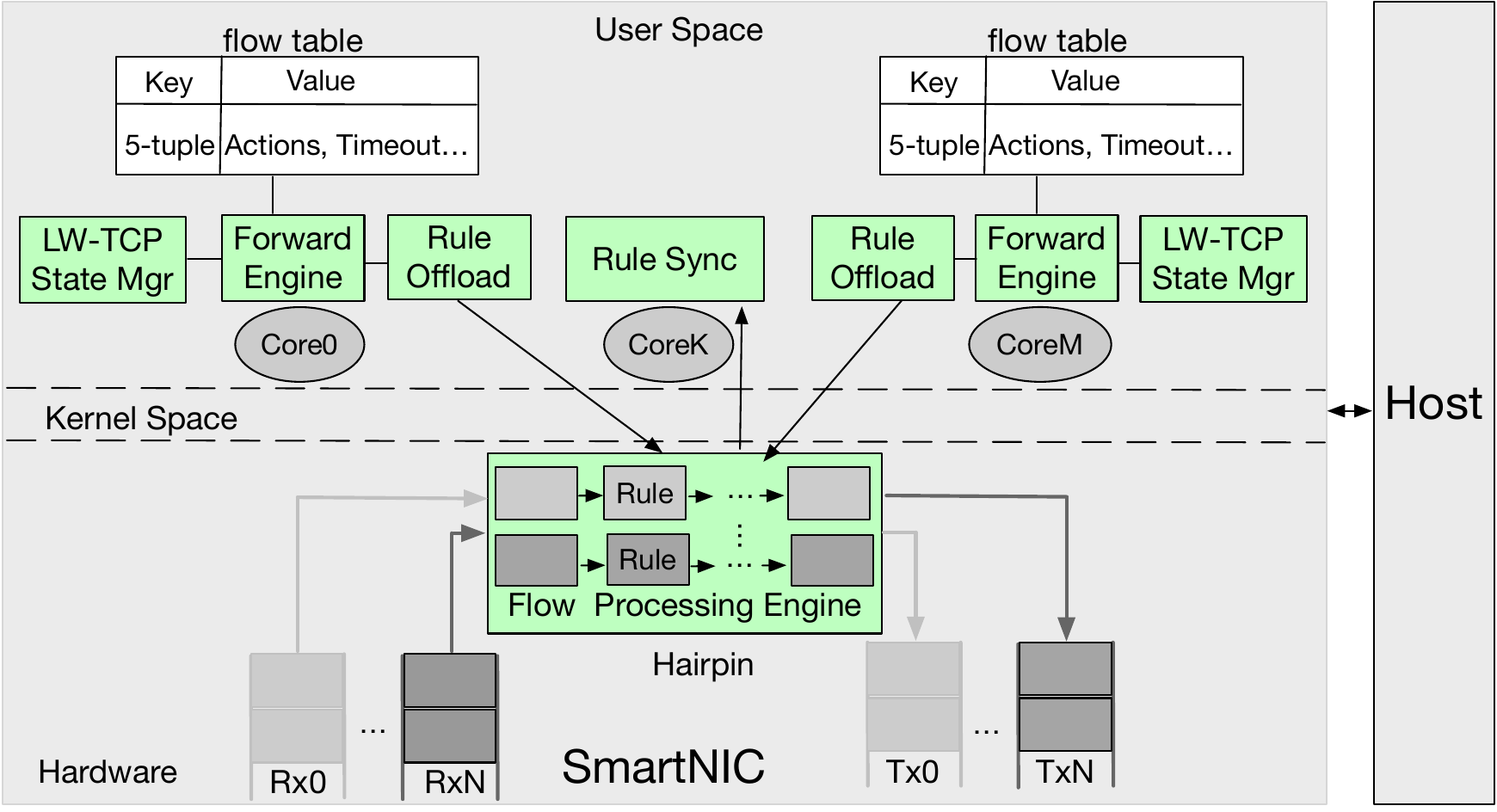}
\caption{\small Laconic Architecture. Laconic steers incoming packets to a SmartNIC core using a flow processing engine. Laconic can use the flow processing engine to offload packet-processing logic, by inserting match-action rules and hairpinning subsequent packets in the flow. The flow table is shared among cores in on-path SmartNICs or distributed across cores in off-path SmartNICs.}
\label{fig:laconic-arch}
\end{figure}

We identify three key techniques to effectively use SmartNICs: lightweight network stack, lightweight synchronization for shared data structures, and hardware-accelerated packet processing. The first two techniques streamline the packet-processing logic performed by the load balancers. The third technique takes advantage of hardware flow processing engines available on SmartNICs for accelerating packet processing and rewriting. Figure~\ref{fig:laconic-arch} shows how \sys integrates the three techniques.

\noindent \textbf{Lightweight network stack:} To avoid the costs associated with the use of a full-featured network stack, we consider an alternate design that utilizes a lightweight packet forwarding stack on the LB and relies on the end-hosts themselves to achieve the desired end-to-end properties. We co-design the application layer load-balancing functions with a streamlined transport layer. Only a subset of the packets have their payloads inspected and modified by the application layer (e.g., client HTTP request packets and HTTP headers on server responses), and the LB's transport layer is responsible for reliable delivery only for this subset. \sys relies on the TCP stacks at the server and the client to provide reliability, sequencing, and congestion control for the remaining packets; it performs simple transformations to the packet headers, without needing access to their payloads, to transfer these responsibilities.


\noindent \textbf{Lightweight synchronization for shared data structures:} As we streamline the processing logic, the synchronization costs for concurrent access to shared data structures would limit performance.  These synchronization costs are incurred when different cores on a SmartNIC handle a flow's packets. (On-path SmartNICs eschew using RSS to spread packets across cores evenly. Even when RSS is available, say on off-path SmartNICs, the packets associated with the two different flow directions of a given client-server connection will be handled on different SmartNIC cores.) We, therefore, design highly concurrent connection table management mechanisms for load balancers. We use the semantics of the shared data and the context in which the data structures are accessed to optimize the concurrency control. For instance, some elements of the connection table are initialized during connection setup and never mutated afterward. Other elements, like the expiry timestamp for connection table entries, can benefit from relaxed update semantics.

\noindent \textbf{Acceleration with Hardware Engine:}  Our design considers the flow processing engine on off-path SmartNICs as a packet-processing accelerator on which we can offload the load-balancing logic. This allows us to perform the common-case packet rewriting logic on the hardware accelerators, thus reducing the packet processing burden on the generic ARM cores. However, these hardware accelerators cannot be used for all flows, as adding or removing match-action rules incurs moderate latency and is throughput limited. \sys uses hardware acceleration only for sufficiently large flows. \sys also relies on the general-purpose cores to perform packet rewriting while rules are being added and hides the latency of rule deletions with the connection establishment phases of subsequent flows.




\subsection{Lightweight Network Stack}
\label{light-networking-stack}

\subdraft{

\begin{itemize}
    \item Our goal of the project:
    \item keep the system fast
    \item exam what is needed and what is not needed: two functionality, header modification, and header examination
    \item In a HTTP flow, we typically only need to exam the HTTP header
    \item HTTP header typically is shorter, just a few packets
    \item As a result, not the entire flow needs reconstruct
\end{itemize}
\begin{itemize}
    \item remove: shift congestion control to the endpoint
    \item half: reliable delivery only kept for the inserted portion
\end{itemize}
}


\begin{figure}[tbp]
\centering
\includegraphics[width=0.5\textwidth, keepaspectratio]{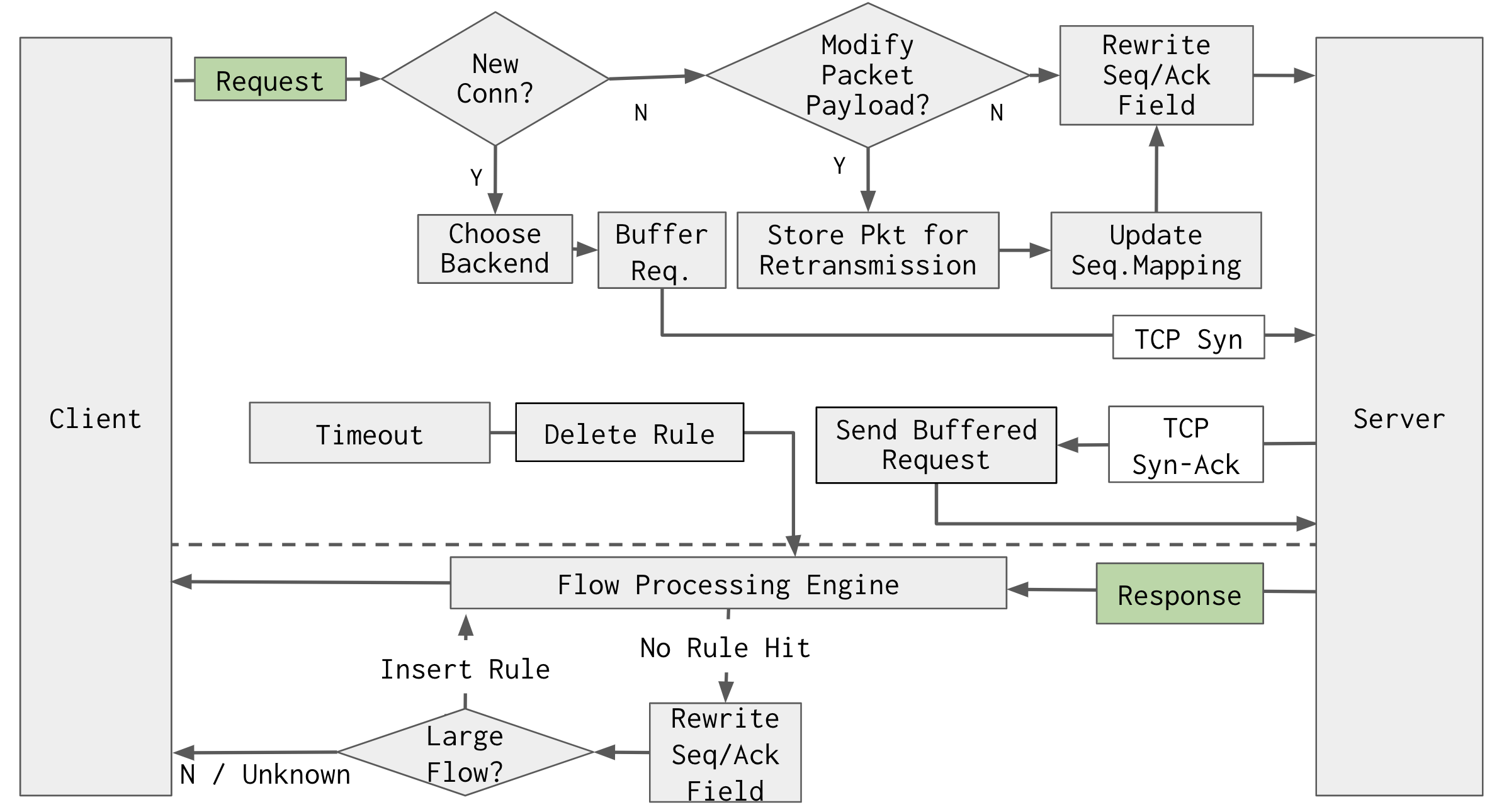}
\caption{\small Workflow of Laconic lightweight network stack. The upper part shows the path from the client side to the server side and the lower part vice versa. To simplify our presentation without losing generality, we assume the HTTP header processing and modification only happen from the client to the server side. The flow processing engine is a hardware engine that has the ability to handle packets entirely, bypassing the cores on the SmartNIC.}
\label{fig:laconic-netstack}
\end{figure}
Our description focuses on two crucial functionalities of L7 load balancers, as these motivate the  processing performed by the load balancer and its associated state.


\vspace{0.1in}
\noindent \emph{Route HTTP requests}.
An L7 load balancer can be configured to route an HTTP request by parsing the fields in the HTTP header, such as the URL prefix.  L7 load balancers buffer the entire HTTP header and then identify the backend using configured match rules.

\vspace{0.1in}
\noindent \emph{Modify HTTP headers}. L7 load balancers update HTTP header requests and responses by adding specific header values. For example, headers protect against XSS or CSRF attacks. Another common usage is to insert the client's IP address into the request header or a server identifier into a response cookie to enable a consistent level of service per client.

\sys's approach to packet processing is designed to be lightweight. Laconic utilizes a simple forwarding agent that constructs only the necessary packet content to make routing decisions and only buffers modified packet content. In particular, \sys\ buffers and processes packets from the client that corresponds to the HTTP requests identifies the appropriate backend servers, and ensures the reliable delivery of these (potentially modified) packets to the server. For packets sent by the server to the client, \sys\ performs some simple packet rewriting using a limited amount of state. Crucially, Laconic depends on the client to keep track of which packets it has received, and it modifies ACKs from the client to inform the server about which packets need to be retransmitted. \sys\ thus does not maintain any state for these packets, and it relies on client/server TCP logic for reliable end-to-end delivery and congestion control. Figure~\ref{fig:laconic-netstack} shows the overall packet flow of our network stack. 

\subsubsection{State maintained}
\label{subsec:designstate}
Each table entry in the connection table maintains a state machine representation of a connection.  The connection state is one of the following. \newline \texttt{FRONT\-\_ESTABLISHED}: the client has created a TCP connection with the load balancer, but a backend has not been determined; \verb|SYN_SENT|: The backend has been identified, but the connection hasn't yet been fully set up; and \verb|ESTABLISHED|: the connection to the backend has been set up.

\sys also maintains a shallow buffer for each connection to buffer the packets received before the backend connection is established. The forwarding agent requires this buffer to construct the HTTP headers and parse information, such as the URL associated with the request.

At its core, the forwarding agent bridges two TCP connections and simply relays packets after appropriately rewriting the packet and ACK sequence numbers. The forwarding agent has to maintain the mapping between the sequence number spaces of the two connections.  Since the load balancer can insert new header fields that will change the sequence number mappings, we maintain an array of \textit{insertion points}. Each insertion point records two pieces of information: the data offset where content is inserted and the size of the inserted data. For each packet, the load balancer performs a linear scan through the array, computes the total amount of inserted data before the packet, and uses this size value as an offset to adjust the sequence and ACK numbers.

\begin{figure}[tbp]
\centering
\includegraphics[width=0.35\textwidth, keepaspectratio]{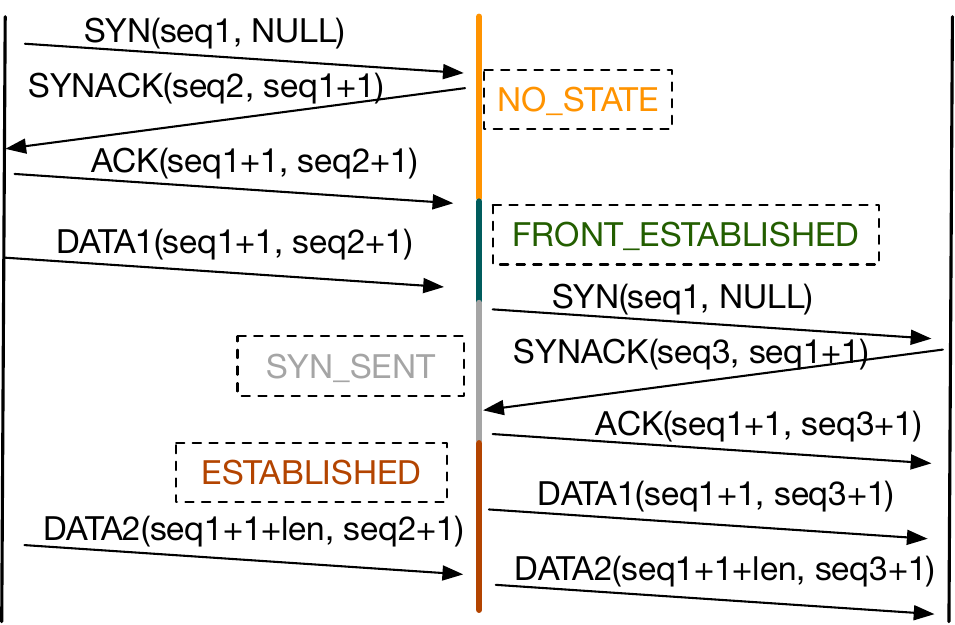}
\caption{Connection setup flow chart}
\label{fig:l7pktsetup}
\end{figure}

\subsubsection{Connection setup}
\label{l7conn:setup}

Figure~\ref{fig:l7pktsetup} demonstrates the entire connection establishment workflow. 

\vspace{0.05in}
\noindent \emph{Client's SYN received:} The load balancer sends an SYNACK packet with a sequence number chosen according to the SYN cookie and the same TCP options as backend servers.


\vspace{0.05in}
\noindent \emph{Client data received:}    
\sys buffers the packets received from the client till it can determine a backend. In particular, the load balancer will buffer client packets until it can reconstruct and parse the header fields, e.g., the hostname and the request URL. 
A connection table entry is created when the first client packet with payload is received. 

\vspace{0.05in}
\noindent \emph{Backend connection setup:} After the load balancer has received sufficient client data (e.g., \texttt{DATA1} in Figure~\ref{fig:l7pktsetup}), it can determine the backend. It then sends an SYN to the backend and completes the three-way handshake. \sys records the sequence and ACK numbers in the connection table and then forwards the buffered client packets to the backend server, possibly after modifying any desired header. If the headers were modified, the forwarding agent holds on to the buffers until it receives the ACKs from the server; or else it releases them immediately.  In the latter case, it will pass along duplicated ACKs to the client, which will then retransmit the data.

\vspace{0.05in}
\noindent \emph{Relay established:} From this point, the connections to the client and the backend server are established and bridged. The subsequent packets will be forwarded directly without any buffering, as we will discuss next.

\subsubsection{Packet processing}
\label{subsubsec:pktprocessing}


We discuss how the forwarding agent relays packets by appropriately modifying the sequence and ACK numbers.  With \verb|HTTP 1.1|~\cite{fielding1999rfc2616}, a persistent connection can convey multiple HTTP requests over the same TCP connection. As a result, content insertion or modification at different points of a TCP flow  is required to support the modification of multiple requests.

\begin{figure}[tbp]
\centering
\includegraphics[width=0.4\textwidth, keepaspectratio]{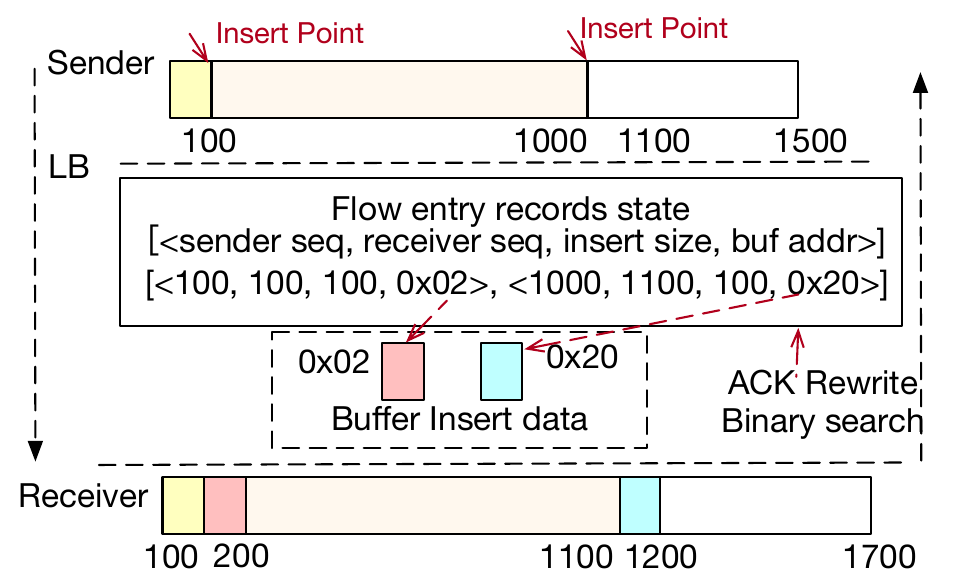}
\caption{An example of content insertion. The top bar of the figure shows the original sequence number sent by the client. \sys{} adds 100 bytes (red) at the original sequence number 100 and another 100 bytes header (blue) at the original sequence 1000. The sequence seen by the server is shown at the bottom of the figure.}
\label{fig:inserteg}
\end{figure}

Figure~\ref{fig:inserteg} shows an example of a TCP connection handled by the load balancer. The sender sends a sequence of requests over multiple packets resulting in a flow of size 1500 bytes to the receiver. \sys determines two locations where we need to perform content insertions. For simplicity, we designate the sequence number space starting from zero. The insertion locations are at 100B and 1000B (where "B" stands for bytes) when viewed from the sender side and at 100B and 1100B when viewed from the receiver.

When the load balancer performs an insert, it records the insertion point's sequence number in the flow's entire sequence space, as viewed by the sender and the receiver.
Further, it splits the original packet at the insertion point and transmits the original packet fragments and the inserted content as separate packets.

\sys appropriately rewrites the ACK numbers in the presence of multiple insertions and handles retransmissions of inserted data.
There are several cases to consider. If the receiver's ACK is far beyond the inserted location (i.e., if the ACK number is higher than $1200+MTU$ in our example), we simply use the last offset stored in the connection table and rewrite the ACK using this offset. If the ACK is between two insertion points (i.e., it is in the range of $(201+MTU, 1100)$ in our example), we use the offset of the previous insertion point to calculate the ACK sent to the sender. 
If the ACK number is exactly at an insertion point and if multiple such duplicate ACKs have been received, then the load balancer retransmits the inserted content.  The last case is that the ACK number is exactly after an insertion point (e.g., 201 or 1201 in our example), in which case we suppress the ACK instead of relaying it and triggering duplicate ACK processing on the sender. In the case of packets containing SACKs (selective acknowledgments), \sys performs the ACK number transformations on every SACK block.

ACK packets are also used as the signal to garbage collect all the buffers buffered at the load balancer. The load balancer will check the ACKs against the stored buffers and release those that have been ACKed.

Note that the ACK number transformations ensure that the sender can use duplicate ACKs to detect and retransmit lost packets.  The forwarding agent is responsible for the reliable delivery of only the inserted content, with the sender responsible for the reliability of all other content. 
\subsection{Lightweight synchronization for shared data}
\label{subsec:design-sync}

We now present the design of efficient synchronization for the load balancer's data structures, e.g., the connection table. Ideally, we would employ a scheme such as receive-side scaling (RSS), which would allow each NIC core to have exclusive and lock-free access to its shard and avoid sharing data structures across NIC cores. This is hard in the context of both L4 and L7 LBs, as the connection table state would be accessed by traffic from both directions, and RSS would inevitably map the forward and reverse directions to different cores. Thus, \sys's connection table needs to provide good scalability with concurrent accesses of per-flow states from multiple NIC cores. 

The load balancer records the states of each flow in a connection table, which is used to select a backend server and update the IP and port attributes in a packet header. At a high level, the flow state recorded by \sys in the connection table could be categorized into three kinds:

    $\bullet$ Decision of the backend server assignment.
    
     $\bullet$ Timeout information of connection.
     
     $\bullet$ TCP state and seq/ack number mapping.

For each category of state, \sys uses different strategies in the design of the connection table to accommodate its access pattern.

\emph{Backend Assignment:} Once the backend is chosen for a specific client connection, the assignment will stick to the connection through its lifetime, meaning that the vast majority of accesses to this state would be read-only. Therefore, \sys chooses to design its connection table based on Cuckoo hashing~\cite{li2014algorithmic}, which provides predictable and constant-time lookups. Cuckoo hashing achieves this property by limiting an entry's position to two hashed locations within the table, but it might have to periodically shuffle the entries to accommodate new insertions. We now discuss how we can eliminate the use of locks during the lookup and still achieve correctness.
For read operations, \sys uses a version-counter-based approach to prevent the lookup operations from reading concurrent write operations: before and after \sys reads the hash table slot with a matching key, a version counter is read from the table entry. A mismatched version number indicates that a concurrent write has been performed on the entry, in which case \sys will retry the read operation. With this approach, no lock is held during a lookup operation to the connection table. To further improve concurrency, \sys's connection table adopts the optimistic locking scheme in \cite{li2014algorithmic} to reduce the critical section during entry shuffles triggered by inserts and minimize the amount of time a write operation needs to hold locks.

\emph{Flow Timeout: } Tracking the last access time of each flow is important for \sys to correctly time out and remove entries of old flows and reclaim its resources. However, recording the access time for the per-flow state could impact scalability. This is because the entry needs to be updated for every access, effectively making every single access a concurrent update, typically performed while holding a lock. To mitigate the potential inefficiencies caused by such implicit writes, \sys uses a ``blind write'' to update the timestamp during the lookup operation. If a later version number check reveals that the record has been updated and is now storing state for a different flow, \sys will not attempt to revert the update and simply retries the timestamp update. With this approach, \sys could spuriously update the TTL of a different record, thereby delaying its garbage collection, but it also ensures that a received packet would always bump up the corresponding flow's TTL.

\emph{TCP state and seq/ack number mapping: } For L7 load balancing, \sys needs to maintain the TCP state (LISTEN, SYN-SENT, ESTABLISHED, etc.) and the mapping of seq/ack number between the client side and server side. However, with the help of RSS, state transitions of the TCP state machine of a single connection will always be performed on the same NIC core. Therefore, \sys updates the state machine without acquiring any locks. We observe that the access of the seq/ack number mapping follows a consumer/producer pattern, so we use lock-free mechanisms for appending to and reading from this list.

\subsection{Acceleration with Hardware Engine}
\label{subsec:hardwareoffloading}


\subsubsection{Capabilities of Flow Processing Engine}

On off-path SmartNICs, the hardware flow processing engine can significantly accelerate packet processing. Conceptually, it works similarly as a switch with the Reconfigurable Match-Action Table (RMT) architecture. Multiple match-action tables can exist in the hardware, and match-action rules for various packet fields of a packet can be inserted dynamically. We now describe how \sys\ can take advantage of these capabilities. 
We describe two key constructs available in flow processing engines that together aid in offloading packet-processing logic.

\noindent \textbf{On-NIC rules:} 
For NICs equipped with the flow processing engine, DPDK provides a \texttt{RTE\_FLOW} interface to insert and delete rules into the NIC. These rules have two fields: a match field and an action field. 

For the match field, the application can specify fields from packet headers to match against, such as IP/MAC address and TCP/UDP port numbers. This allows the application to match packets with specific values for those fields.

For the action field, applications can specify two types. The first type is \emph{fate actions}, which determines the packet's final destination, such as dropping it or moving it to a different queue. The second type includes \emph{non-fate actions} such as packet counting and rewriting headers, including TCP port, IP address, and sequence numbers. Header updates can change a specific field or add a predetermined constant value. The flow processing engine can also set an expiry time for a rule and notify the application if no packet matches the rule after a certain time.

\noindent \textbf{Hairpin queue:} Hairpin queue is a feature first introduced in \texttt{DPDK 19.11} \cite{DPDKhairpin}. It is similar to a loopback interface, but it operates on the network-facing ports of the NIC. The hairpin queues are a pair of connected RX and TX queues.


\noindent \textbf{Benefits:} With a combination of on-NIC rules and hairpin queues, we can dramatically reduce the involvement of general-purpose cores in the load-balancing datapath. Specifically, if we were to insert rules to match packets with known flows, specify the appropriate actions to rewrite the packet contents (e.g., the sequence number and ACK number fields), and send it out through a hairpin queue, then most of the packet-processing logic of our network stack can be offloaded to the NIC flow processing engine. Furthermore, latency can be improved since all the forwarding data path is inside the NIC pipeline and bypass computing cores.

\noindent \textbf{Challenges:} However, there are two types of challenges for utilizing the flow processing engine. 

The first type is a functional challenge: current SmartNICs do not support a range match operation, which would be useful in the context of \sys\ for matching a range of sequence numbers. And the match-action rule can not be performed on the selective acknowledgment (SACK) blocks in the case where a receiver is signaling holes in the received sequence number space.

The second type is a performance-related challenge: rule insertions and deletions can both be expensive in terms of latency. On the BlueField-2, the cost of a non-blocking rule insertion with batching is 25 microseconds in our tests, whereas the cost of a blocking rule insertion is substantially higher at 305 microseconds.
These measurements indicate that rule updates should be performed sparingly, especially in the context of short flows, and that rule insertion costs should be overlapped with other tasks, through the use of non-blocking rule insertions.

\subsubsection{Using Flow Processing Engine}
We now describe how \sys{} uses the flow processing engines. 

\vspace{0.05in}
\noindent \textbf{What functionalities to offload?} We first note that the flow processing engines do not have advanced parsing capabilities. Therefore, request processing logic that operates on content sent from the client to the server (e.g., parsing HTTP headers, requests, etc.) is not offloaded but rather performed on a SmartNIC ARM core.

Second, given the overheads of performing rule insertions and deletions and the latency cost associated with performing the rule update, we do not perform offloads for small flows. The HTTP parsing logic on the ARM core identifies the resource length (specified as part of the HTTP headers) and offloads packet rewriting logic only for moderate- to large-sized flows. We simply calculate  the threshold of flow size as follows. We assume the response size is $B$, and the TCP MSS size is $MSS$. Let the required time to process a single packet on the lightweight network stack fast path is $T$. The \texttt{RTE\_FLOW} insertion and deletion times are $P$. In order to have a benefit of offloading, $\frac{B}{MSS}T \ge P$, so $B \ge \frac{P}{T}MSS$.

Finally, for these offloaded flows, packet rewriting is performed only on packets sent from the server to the client; packets sent from the client to the server, which is a small portion of the traffic load, are handled on the SmartNIC ARM cores. We made this choice based on the limitation that Bluefield-2 cannot rewrite the SACK option in TCP headers. We also observed by experiment that disabling SACK incurs a dramatic loss in TCP flow performance. 

\vspace{0.1in}
\noindent \textbf{When  to offload?} We now describe the life cycle of flow rules from insertion to deletion.  Once a server response exceeds the size threshold, a general-purpose core inserts a rule to match and modify packets sent from the server to the client. This rule matches the TCP 5-tuple corresponding to the server-to-load balancer connection, with the action of updating the sequence and ACK numbers associated with the packet using the correct offsets. This rule insertion is performed using a non-blocking operation so as to reduce the overhead associated with performing the operation. Due to the non-blocking operation, subsequent packets from the server could still be sent to a generic ARM core if the insertion hasn't been completed, but correctness is ensured as both the ARM core and the flow processing engine are performing the same operation.

To delete the rules, the SmartNIC cores continuously check the client ACKs as to whether the entire server response has been received or not. ACK packets from clients are continued to be handled by ARM cores instead of the flow processing engines. If the entire response has been received, then the SmartNIC core initiates the process of removing the packet update rule in preparation for having a clean connection for the next client request. The rule deletion operations are first conveyed to a dedicated core, which batches outstanding rule deletions and then performs them synchronously. By delegating this to a dedicated thread and batching such operations, we reduce the overhead of blocking rule updates. Finally, when the next client request is received, the receiving core first checks (and possibly waits) for the condition that the prior rule associated with the connection has been removed before processing the client request.


\noindent \textbf{Achieve RSS with the engine:}
We implement a custom RSS using the engine, which greatly simplified the need for handling concurrent access to shared data structures from different cores. Traditional RSS that operates on a flow's 5-tuple is insufficient as the client to LB, and the server to LB segments correspond to different 5-tuples, and RSS would steer the packets received from the two directions into two different cores. Thus in Section~\ref{subsec:design-sync}, we propose lightweight synchronization for shared data. We can eliminate the synchronization  mechanisms if packets from a given client-server connection are processed by the same core in either direction of the packet flow. 


Specifically, we pre-install a set of rules to steer incoming packets to cores solely based on the port numbers in the TCP header. We partition the port number space into multiple shards and assign each shard to a processing core. We insert a rule into the flow processing engine for each allocated TCP port number so that client packets with the source port number and server packets with the destination port number are sent to the assigned core. This ensures that packets from the same client-server connection are always processed by the same core. By doing so, we can eliminate the need for synchronization in accessing the flow table, and each core can have its own shard of the connection table. Note that this method is only needed for short flows whose packet processing is not offloaded to the engine, which also helps avoid load imbalance issues that may occur when processing large elephant flows using the same NIC core.

%% file: sections/eval_v2.tex
\section{Evaluation}
\label{sec:eval}
In this section, we conduct both microbenchmarks and end-to-end application benchmarks to understand the performance of \sys. Specifically, we aim to answer the following three questions:

$\bullet$ Can \sys's design provide end-to-end performance benefis compared with existing load balancers?

$\bullet$ What benefits are provided by each of the individual techniques used in Laconic's design?

$\bullet$ How well does Laconic perform under real-world application workloads?



\subsection{Experiment Setup}
\textbf{Implementation:} We implement \sys on both on-path and off-path SmartNICs with \texttt{DPDK 21.11}. We implement \sys in about 6k lines of C++ code. 

\begin{figure}[htbp]
\centering
\includegraphics[width=0.45\textwidth, keepaspectratio]{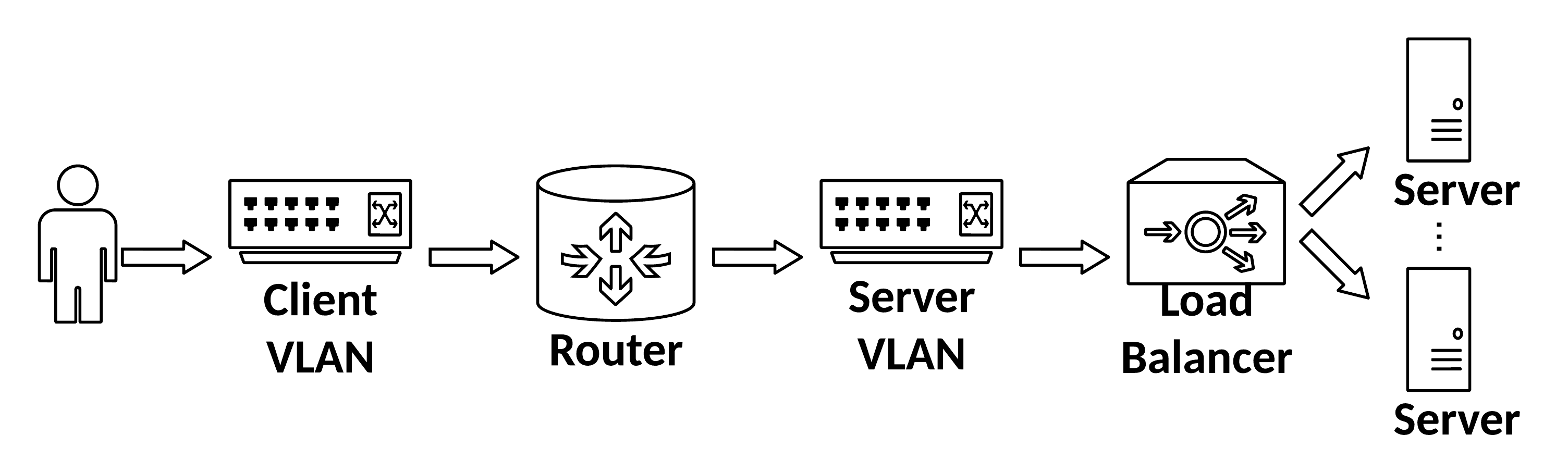}
\caption{Workflow for a request from the client to the server.}
\label{fig:eval-topo}
\end{figure}

\textbf{Testbed:} Our testbed consists of five x86 servers with Intel Xeon 5218 CPU and 64GB of memory. Each server is equipped with a Mellanox ConnectX-5 two-port NIC. The link speed of each individual port is 100~Gbps. We use a dedicated set of servers to host SmartNICs. The on-path SmartNIC we use is a Marvell LiquidIO 3 CN3380 with 2x50~Gbps ports. The off-path SmartNIC used in our deployment is an Nvidia BlueField-2 MBF2H516C-CECOT with 2x100~Gbps ports. An Arista DCS-7060CX switch is configured with VLAN and L4 routing. We partition the clients and servers into two different subnets,  which is a common setting in actual deployment. We allocate two of the five servers as clients and the other three act as servers. The load balancer operates within the server subnet and is responsible for receiving and forwarding requests through the same NIC port. This setup, referred to as "router-on-a-stick," allows the load balancer to manage the incoming requests and distribute the requests among the backend servers. Figure~\ref{fig:eval-topo} shows the path of a request to one of the backend servers. 
\begin{figure*}[ht!]
\centering
    \begin{subfigure}[a]{0.23\textwidth}{
                    \includegraphics[width=1\textwidth]{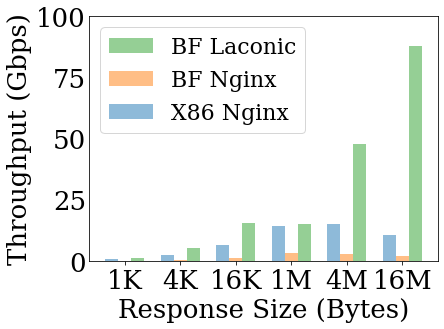}}
            \caption{BF-2 with Single Core}
            \label{fig:singlecorebfgbps}
    \end{subfigure}
    \begin{subfigure}[a]{0.24\textwidth}{
                    \includegraphics[width=1\textwidth]{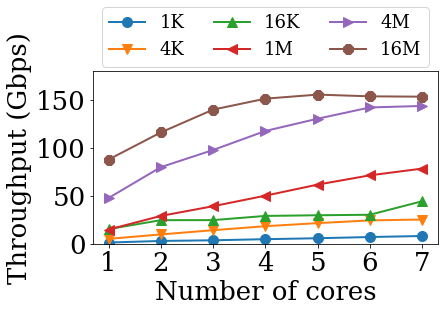}}
            \caption{BF-2 Core Scalability}
            \label{fig:bfdiffcore}
    \end{subfigure}
    \begin{subfigure}[a]{0.23\textwidth}{
    \includegraphics[width=1\textwidth]{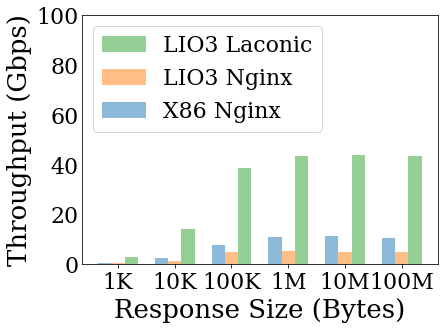}}
                
            \caption{LIO-3 with Single Core}
             \label{fig:singlecorelio3gbps}
    \end{subfigure}
    \begin{subfigure}[a]{0.24\textwidth}{
                 \includegraphics[width=1\textwidth]{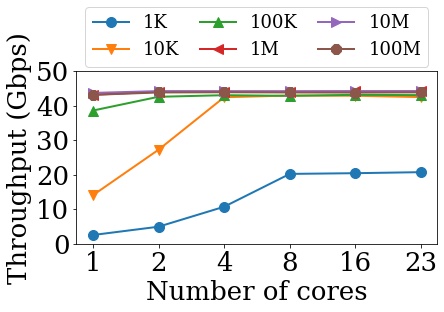}}
            \caption{LIO-3 Core Scalability}
            \label{fig:lio3diffcore}
    \end{subfigure}
\caption{Throughput of \sys on BlueField-2.}
\label{fig:throughput}
\end{figure*}

\begin{figure*}[ht]
\centering
    \begin{subfigure}[a]{0.24\textwidth}{
                \includegraphics[width=1\textwidth]{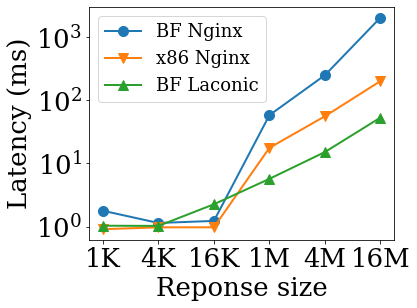}}
        \caption{BF-2 with Single Core}
        \label{fig::latencysinglecore}
    \end{subfigure}
    \begin{subfigure}[a]{0.24\textwidth}{
                \includegraphics[width=1\textwidth]{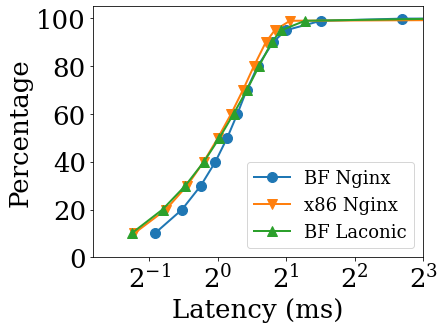}}
            \caption{1KB with BF-2 Single Core}
            \label{fig:latency1Kcdf}
    \end{subfigure}
    \begin{subfigure}[a]{0.24\textwidth}{
                \includegraphics[width=1\textwidth]{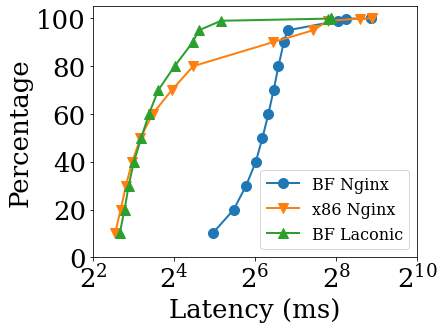}}
            \caption{4MB with BF-2 Single Core}
            \label{fig:latency4Mcdf}
    \end{subfigure}
    \begin{subfigure}[a]{0.24\textwidth}{
                \includegraphics[width=1\textwidth]{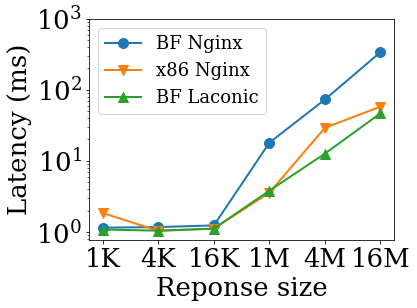}}
            \caption{BF-2 with All Cores }
             \label{fig::latencyallcore}
    \end{subfigure}
\caption{Latency of \sys on BlueField-2}
\label{fig:latencyrespsize}
\end{figure*}

\textbf{Baseline:} There are multiple widely adopted L7 load balancers on the market, including Envoy~\cite{envoy}, Nginx~\cite{nginx}, HAProxy~\cite{haproxy}, etc. We choose Nginx as our baseline since it is widely adopted in practical deployment and provides higher performance than others~\cite{dropboxnginx}.

\textbf{Client and backend server:} \texttt{wrk} and \texttt{wrk2} are used with customized Lua scripts as the client software. We generate a list of static files with various file sizes to responding the HTTP requests of \texttt{wrk}. We use the metric of requests per second (RPS) reported by \texttt{wrk}, and we multiply it by the file size to get the goodput measurement. Nginx is configured as the backend server software which responds to \texttt{wrk} requests with static files of varied file sizes. 
\subsection{End-to-end Throughput}
\label{sec:throuhghput}

We evaluate the throughput performance of Laconic on off-path BlueField-2 SmartNIC and on-path LiquidIO3 SmartNIC (Figure~\ref{fig:throughput}) while varying the number of cores and file sizes of wrk requests.

\textbf{Using a single core:} To show the processing capability of \sys, we first evaluate the throughput on just a single core by varying the response size. Figure~\ref{fig:singlecorebfgbps} clearly shows that the single-core performance of \sys on BlueField-2 outperforms Nginx on x86. In this experiment, we set the threshold of offloading onto the flow processing engine as 1MB. With the help of the hardware flow processing engine, for response sizes of 16M, \sys achieves 8.7x throughput compared with Nginx x86. As shown in Figure~\ref{fig:singlecorelio3gbps}, \sys on LIO3 performs much better than Nginx, even though Nginx runs on a more powerful x86 core. Although LiquidIO3 does not have a hardware flow processing engine, the throughput of \sys is still 4.5x better compared with x86 Nginx.

\textbf{Scaling to more cores}: To show the scalability behavior of using multiple cores, we measure the throughput by varying the number of cores. Figure ~\ref{fig:bfdiffcore} shows Laconic performance on BlueField-2. While offloading the processing into a hardware engine for messages larger than 1M, we observe that throughput scales with more cores for processing large messages, which can achieve up to 150 Gbps throughput. For small messages, we don't offload processing onto the flow processing engine. Throughput for processing small messages also can scale with more cores. In Section \ref{sec::benchmarkthreedesign}, we will explain more about the scalability of processing small messages and why the flow processing engine has a limited impact on its scaling.
Figure~\ref{fig:lio3diffcore} depicts the scalability performance for LIO3, where we see linear increases in throughput till the line rate is achieved, except for the 1KB small message flow case due to frequent connection setup and tear-down. In Appendix \ref{appendix::throughputallcore}, we show zoomed-in figures of the throughput performance with all cores.

\begin{figure*}[ht]
\centering
    \begin{subfigure}[a]{0.24\textwidth}{
                 \includegraphics[width=1\textwidth]{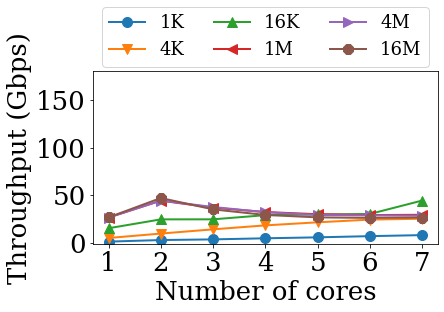}}
        \caption{Non-Offload}
        \label{fig:throughputnonoffload}
    \end{subfigure}
    \begin{subfigure}[a]{0.24\textwidth}{
                         \includegraphics[width=1\textwidth]{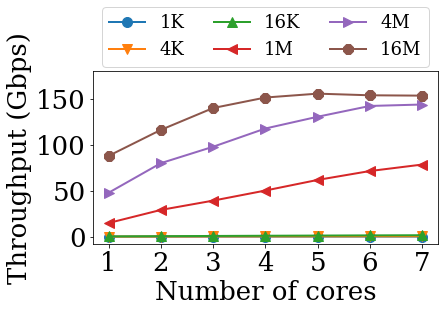}}
            \caption{All-Offload}
            \label{fig:throughputallffload}
    \end{subfigure}
    \begin{subfigure}[a]{0.24\textwidth}{

    \includegraphics[width=0.97\textwidth]{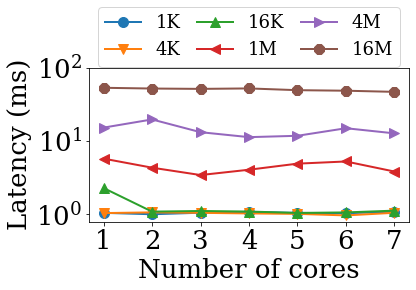}}
    
            \caption{Offload Msg. >1M}
            \label{fig:latencyoffload}
    \end{subfigure}
    \begin{subfigure}[a]{0.24\textwidth}{
                 \includegraphics[width=0.97\textwidth]{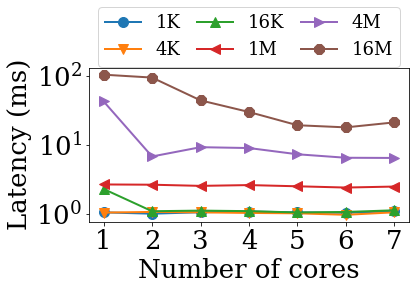}}
            \caption{Non-Offload}
            \label{fig:latencynooffload}
    \end{subfigure}
    %
\caption{Microbenchmark on Offloading to Flow Processing Engine}
\label{fig:offloadengine}
\end{figure*}

\subsection{End-to-end Latency}

We next study Laconic performance in terms of latency as Figure~\ref{fig:latencyrespsize} shows. We use \texttt{wrk2} with setting a target throughput as 50\% peak throughput measured in the previous experiments. In this experiment, we focus on the Laconic on BlueField-2 because Laconic on BlueField-2 achieves higher throughput performance than LiquidIO-3 according to the throughput experiments in Section \ref{sec:throuhghput}. 

\textbf{Using a single core:} Figure~\ref{fig::latencysinglecore} shows average latency results  with a single core. Despite our design running on BlueField-2 with higher throughput, Laconic's latency is comparable to that of the Nginx running over x86 servers. Moreover, Laconic can achieve lower latency for processing large message sizes compared with other baselines by efficiently offloading onto the flow processing engine.

As Figure~\ref{fig:latency1Kcdf} and Figure~\ref{fig:latency4Mcdf} show, we further zoom in to specific response sizes to plot the cumulative distribution function (CDF) of latencies for all requests. \sys  on BlueField-2 achieves lower latency when compared to Nginx running on the ARM cores or the x86 cores for both short and large response sizes, which benefit from our lightweight network stack and efficient synchronization mechanisms. Additionally, large messages are processed by the flow processing engine without much involvement of cores. 

\textbf{Scaling to all cores:} Figure~\ref{fig::latencyallcore} shows the average latency while Laconic uses all ARM cores on BlueField-2. Compared with using only one core, \sys achieves lower latency with more cores. Also, in the multi-core scenario, \sys still outperforms other baselines. In Section \ref{sec::benchmarkthreedesign}, we also investigate how the core scalability affects the latency by varying the number of cores.

\subsection{Evaluating the Benefits of Key Techniques}
\label{sec::benchmarkthreedesign}

\textbf{Offloading to Flow Processing Engine}: Figure~\ref{fig:offloadengine} 
shows the benefits of using a hardware flow processing engine. In this experiment, we vary the offloading threshold such that requests with response size larger than the threshold were processed by the flow processing engine, while ARM cores were used for requests with response sizes smaller than the threshold. Note that prior to fully offloading the flow processing, the initial packets are processed by the ARM cores for establishing connections and inserting flow rules.

Figure~\ref{fig:throughputnonoffload} and ~\ref{fig:throughputallffload} show how offloading affect the throughput. As Figure~\ref{fig:throughputnonoffload} shows, if we use ARM cores to process requests without offloading, it is hard to achieve high throughput even with more cores. There are two reasons for the poor scalability: 1) Memory bandwidth for SmartNIC accessing the on-board memory is limited, which is also reported in other work \cite{iotcp}. 2) For BlueField-2, every two cores share 1MB L2 cache, which could result in cache contention while using multiple cores \cite{bluefield}. Thus, the poor scalability necessitates the offloading to the flow processing engine. Figure~\ref{fig:throughputallffload} shows throughput performance while we process all sizes of requests by offloading to the flow processing engine. We find that the processing of large responses can be scaled with more cores. Because packet processing is offloaded to the flow processing engine, the ARM core is just responsible for setting up the connection and installing flow rules. Additionally, we observed that the processing of small responses does not see improvement from the packet processing engine, largely due to the overhead involved in updating flow rules. This overhead will be discussed in detail below.

Table \ref{tab:rulecost} shows the time cost for updating the flow rule in the flow processing engine. Flows can be inserted or deleted from the engine in batches. From Table \ref{tab:rulecost}, we can see that time cost for updating rules is not negligible for small responses, especially for updating singleton rules. The time cost of updating the rule dominates the flow completion time of small requests. This is the reason why offloading the processing of small responses, such as those less than 1KB, does not improve performance. So, in practice, we only offload the processing of large responses (e.g., >1MB) to the flow processing engine. Also, we use batching to amortize the updating cost. The batch size is determined by the flow arrival pattern.
\begin{table}[htbp]
    \small
    \centering
     \begin{tabular}{ c|c|c } 
   
         \textbf{Batch Size} & \textbf{Insert Latency}  & \textbf{Delete Latency} \\ 
        \hline
         1 & 305.40 us & 57.49 us\\

         2 & 100.48 us & 24.48 us\\
  
         8 & 38.72 us & 19.42 us\\

         16 & 25.39 us & 18.08 us\\

        \end{tabular}
        \caption{Time Cost of Update Flow Rules}
        \label{tab:rulecost}
\end{table}

Figures~\ref{fig:latencyoffload} and \ref{fig:latencynooffload} show how offloading affects the latency. In Figure~\ref{fig:latencyoffload}, we offload processing responses whose size is larger than 1MB. As expected, the latency of processing responses remains stable as we increase the number of cores because, after offloading, ARM cores are not used to process response packets. 
Figure~\ref{fig:latencynooffload} depicts the case where ARM cores process responses without offloading. In this case, it is shown that not offloading to the flow processing engine leads to lower latency when using all cores and similar performance to offloaded processing when using a single core, due to avoiding the rule update costs.

\begin{minipage}{0.48\textwidth}
    \begin{minipage}[t]{0.4\textwidth}
    \small
     \centering
        \begin{tabular}{ c|c } 
    
         \textbf{Msg. Size} & \bf{Mpps} \\ 
          \hline
         1KB & 0.72 \\
   
          4KB & 1.10 \\
      
        16KB & 2.06 \\
        
  1MB & 3.04 \\
     
 4MB & 3.01 \\
       
 16MB & 3.00 \\
     
        \end{tabular}
\makeatletter\def\@captype{table}\makeatother\caption{Throughput}
  \label{tab:netstackpps}
  \end{minipage}
  \begin{minipage}[t]{0.5\textwidth}
  \small
        \begin{tabular}{ c|c } 
        
         \textbf{Top-5 Operation} & \bf{Percent} \\ 
          \hline
         DPDK\_Rx & 40.55\% \\
          
        Parse Resp. Header & 11.60\% \\
    
       Insert Header Value   & 8.01\% \\ 
     
        Reassemble Packet  & 7.36\% \\ 
     
        Lookup Hashtable & 3.56\% \\ 
    
         \textit{Others} & 28.9\% \\ 
      
        \end{tabular}
        \makeatletter\def\@captype{table}\makeatother\caption{Utilization Breakdown}
        \label{breakdown}
   \end{minipage}
\end{minipage}

\textbf{Network Stack:}
We next study the performance of our lightweight network stack. Table \ref{tab:netstackpps} shows the throughput of packet processing with our network stack. Our network stack can achieve up to 3 million packets per second (Mpps) using a single ARM core of BlueField-2. To further  analyze the efficiency of the network stack, we use the \textit{perf} tool to break down the ARM core utilization for operations of packet processing. Among the top-5 operations in terms of ARM core utilization, We find that the DPDK API costs up to 40\%  utilization, which is the internal cost of the DPDK. For other operations that are memory-intensive, they frequently execute memory allocation and population, which are bottlenecked by the memory bandwidth. From the breakdown, we can see that our network stack implementation is efficient, as it is mainly bottlenecked by the DPDK API cost and memory bandwidth.

\begin{figure}[htbp]
\includegraphics[width=0.24\textwidth]{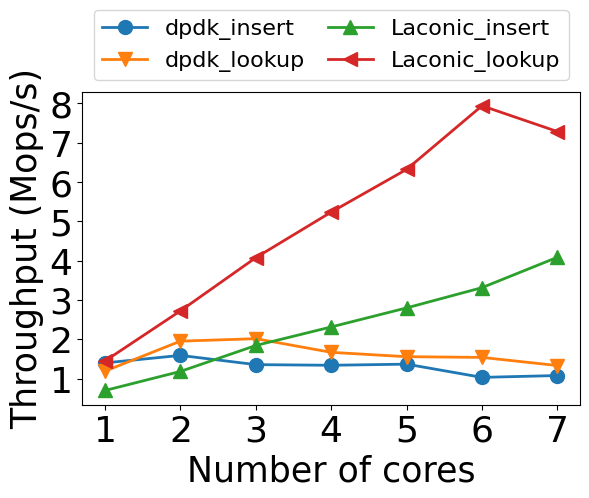}
\caption{Performance of Concurrent Hashtable}
\label{fig::hashperf}
\end{figure}

\textbf{Concurrent Hashtable}: We evaluate our concurrent hash table design. We measure the throughput of two basic hashtable operations, insert and lookup, with the ARM cores. As Figure~\ref{fig::hashperf} shows, the hash table of Laconic can achieve higher throughput than the hash table provided by the DPDK library. Also, the hash table of Laconic scales the performance well with more cores. Further, Table~\ref{tab:netstackpps} shows that Laconic handles 3M packets/sec, a rate that is supported by our hashtable but not the other alternatives.

\subsection{Real-world Workload}
\begin{figure}[ht]
\centering
    \begin{subfigure}[a]{0.23\textwidth}{
                 \includegraphics[width=1\textwidth]{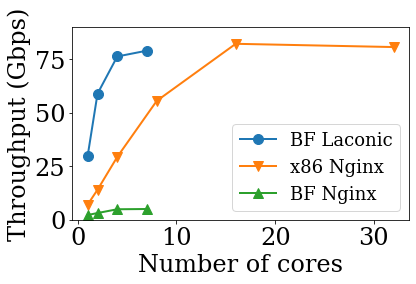}}
            \caption{BF-2 Gbps}
            \label{fig:realgbps-bf}
    \end{subfigure}
    \begin{subfigure}[a]{0.23\textwidth}{
                \includegraphics[width=1\textwidth]{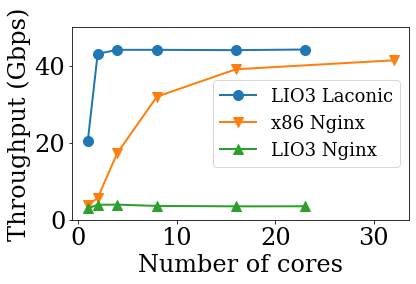}}
        \caption{LIO-3 Gbps}
        \label{fig:realgbps-lio3}
    \end{subfigure}
    %

    %
\caption{Throughput of \sys with Real-world Workload}
\end{figure}

To evaluate the performance of \sys\ under a real-world workload, we adopt the workload from the CONGA work on datacenter traffic load balancing~\cite{conga}. We extract the flow size distribution from their enterprise workload and use it to generate a list of files on the backend server with corresponding sizes. On the client side, we use a customized Lua script for \texttt{wrk} to generate the request according to the distribution. 

Figure ~\ref{fig:realgbps-bf} shows the throughput of \sys with the real-world workload on BlueField-2, which only has eight ARM cores.  \sys can achieve 13.3x throughput with a single core compared to Nginx running on BlueField-2. \sys achieves higher performance with a fewer number of cores up to 75~Gbps; however, Nginx needs more than 16 cores to achieve a similar throughput.

Figure~\ref{fig:realgbps-lio3} shows Laconic performance on LiquidIO3. LIO3 has 24 ARM cores, so we perform the tests on LIO3 with up to 24 cores. We observe that Nginx on LIO3 runs at a very low throughput (5.21~Gbps). Although Nginx on x86 linearly increases the performance, it needs more than 16 cores for just 50G bandwidth. In contrast, \sys only needs up to two ARM cores of LIO3 to match the 50G NIC bandwidth.

%% file: sections/relwk.tex
\section{Related Work}

\textbf{Load Balancers.} There are multiple works investigating high-performance load balancers. Katran~\cite{katran} is an L4 load balancer implemented in eBPF. SilkRoad~\cite{miao2017silkroad} and Hula \cite{katta2016hula}  leverage programmable switches to build a high-performance L4 load balancer. Maglev~\cite{eisenbud2016maglev} proposed by Google takes advantage of consistent hashing. Beamer~\cite{beamer} provides consistency without maintaining state. DRILL \cite{ghorbani2017drill,ghorbani2015micro} performs micro load balancing. Tiara~\cite{tiara} proposes a new hardware architecture for L4 load balancing. All of these works are based on connection-less load balancing and are unable to satisfy the need for connection-oriented L7 load balancing.

\textbf{Network stack acceleration.} TAS~\cite{kaufmann2019tas} provides a host TCP networking acceleration service. FlexTOE~\cite{FlexTOE} and IO-TCP~\cite{iotcp} offload the host TCP stack onto the SmartNICs. Snap~\cite{marty2019snap} is a high-performance networking stack proposed by Google. AccelTCP~\cite{246496} offloads the control path of TCP onto the SmartNIC. Kim, D., Lee, S., and Park, K \cite{kim2020case} explore offloading the TLS control plane onto the SmartNIC. Zilberman \cite{zilberman2022technical} offloads packet processing with XDP.  Netchannel~\cite{netchannel} disaggregates the network stack. They only offload a part of the control path of the networking protocols onto the SmartNIC and do not touch L7 protocols.


\textbf{Measurements for SmartNIC}. Multiple works provide valuable measurement for SmartNICs~\cite{katsikas2021you, liu2021performance}. Girondi et al. measured the performance of different high-speed concurrent hashtables~\cite{girondi2021high}. We consulted these numbers during the design process. 

%% file: sections/conclusion.tex
\section{Conclusion}


We propose Laconic, which explores the potential of offloading L7 load balancing onto SmartNICs and addresses the associated challenges. In Laconic, we employ a lightweight network stack to avoid the costs of a heavy, feature-rich network stack. To minimize the cost of synchronization operations, we design lightweight synchronization mechanisms that enable higher concurrency and scalability. To further reduce the burden on the generic cores on the SmartNIC,  we offload packet processing by effectively utilizing the hardware acceleration engines available on SmartNICs. This paper does not raise any ethical issues.

\clearpage


%% file: sections/appendix.tex
\section{Appendix}

\begin{figure}[htbp]
    \centering
       \includegraphics[width=0.3\textwidth]{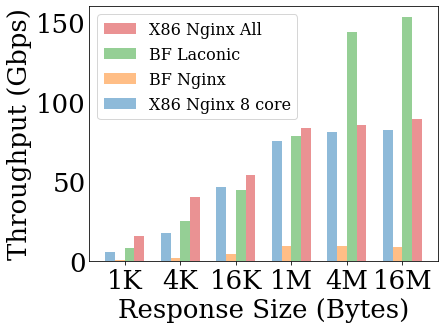}
    \caption{Throughput of \sys on BF-2 with All Cores}
  \label{fig:bffixedgbps}
\end{figure}

\begin{figure}[htbp]
    \centering
      \includegraphics[width=0.3\textwidth]{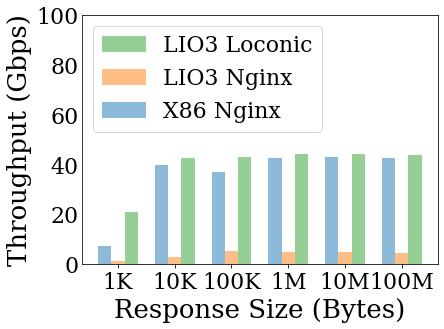}
    \caption{Throughput of \sys on LIO-3 with All Cores}
  \label{fig:allcorelio3gbps}
\end{figure}

\subsection{Throughput of Laconic using all cores}
\label{appendix::throughputallcore}
Figure~\ref{fig:bffixedgbps} shows the throughput of \sys on BlueField-2 with seven ARM cores \footnote{BlueField-2 has eight available ARM cores in total. We reserve one core for the necessary background tasks.}.
The BlueField-2 is equipped with 2x100Gbps ports. We enable the two ports together to further stress the processing power on the ARM cores. For the baseline of Nginx running on the commodity x86 servers, we use 2x100Gbps ports of a ConnectX-5 NIC to make the link speed the same. However, although there are 2x100Gbps ports on ConnectX-5, due to the bottleneck of PCIe 3.0 for testbed servers, only about 120 Gbps can be achieved. As shown in Figure~\ref{fig:bffixedgbps}, Laconic with only seven NIC cores can achieve line rate\footnote{Since the same port on the SmartNIC receives data from both the requests of the clients and also receive the responses from the backend servers, the theoretical upper limit for the load balancer's goodput is lower than the line rate 2x100Gbps.} for large message sizes. For small message sizes, \sys achieves comparable performance with Nginx running on x86. BlueField-2 has wimpy ARM cores, and Nginx running on ARM performs poorly, as expected. There is a clear increase in throughput for messages larger than 1M because \sys offloads the packet processing into the hardware engine for response messages larger than 1M. Other than the first few packets, all the remaining packets in that flow are solely processed by the hardware packet processing engine. This can mitigate the head-of-line blocking problem caused by limited processing power and also relieves the CPU resources for small or future responses. In later Section \ref{sec::benchmarkthreedesign}, we show experiments with varying offloading thresholds.

Figure~\ref{fig:allcorelio3gbps} shows the throughput of \sys on LiquidIO3 50G NIC with 24 ARM cores. Since the LiqudIO3  can only support 50~Gbps link speed, we also cap the x86 baselines at 50~Gbps for a fair comparison of all the experiments running on LiquidIO3. Nginx on the ARM cores of the SmartNIC (bar "LIO3 Nginx") has a low throughput. The throughput plateaus at around 5~Gbps even with 24 cores. \sys can easily reach the full NIC bandwidth with 10K response size. However, Nginx running on an x86 server needs 32 beefy cores to achieve comparable performance. 